\documentclass{article}

    \usepackage[preprint]{neurips_2023}

\usepackage[utf8]{inputenc} 
\usepackage[T1]{fontenc}    
\usepackage[hidelinks]{hyperref}       
\hypersetup{%
  colorlinks=true,
  linkcolor=blue,
  citecolor=blue,
}
\usepackage{url}            
\usepackage{booktabs}       
\usepackage{amsfonts}       
\usepackage{nicefrac}       
\usepackage{microtype}      
\usepackage{xcolor}         
\usepackage{csquotes}

\usepackage{graphicx}

\title{Does Explainable AI Have Moral Value?}

\author{%
  Joshua L. M. Brand\thanks{https://orcid.org/0000-0001-7865-0618}\\
  Institut Polytechnique de Paris, Télécom Paris,\\i3 - Institut interdisciplinaire de l’innovation -- Paris, FR\\
  \texttt{joshua.brand@telecom-paris.fr} \\
  \AND
  Luca Nannini\thanks{https://orcid.org/0000-0002-4733-9760} \\
  $^{1}$Minsait by Indra Sistemas SA -- Madrid, ES\\ $^{2}$Centro Singular de Investigaci\'{o}n en Tecnolox\'{i}as Intelixentes (CiTIUS), \\ Universidade de Santiago de Compostela -- Santiago de Compostela, ES \\
  \texttt{lnannini@minsait.com, l.nannini@usc.es}\\
}

\begin{document}

\maketitle

\begin{abstract}

Explainable AI (XAI) aims to bridge the gap between complex algorithmic systems and human stakeholders. Current discourse often examines XAI in isolation as either a technological tool, user interface, or policy mechanism. This paper proposes a unifying ethical framework grounded in moral duties and the concept of reciprocity. We argue that XAI should be appreciated not merely as a right, but as part of our moral duties that helps sustain a reciprocal relationship between humans affected by AI systems. This is because, we argue, explanations help sustain constitutive symmetry and agency in AI-led decision-making processes. We then assess leading XAI communities and reveal gaps between the ideal of reciprocity and practical feasibility. Machine learning offers useful techniques but overlooks evaluation and adoption challenges. Human-computer interaction provides preliminary insights but oversimplifies organizational contexts. Policies espouse accountability but lack technical nuance. Synthesizing these views exposes barriers to implementable, ethical XAI. Still, positioning XAI as a moral duty transcends rights-based discourse to capture a more robust and complete moral picture. This paper provides an accessible, detailed analysis elucidating the moral value of explainability.
\end{abstract}

\section{Introduction}

In the rapidly evolving landscape of artificial intelligence, Explainable AI (XAI) tools have surfaced as a significant force, seeking to unravel the complexities behind advanced yet opaque artificial intelligence mechanisms. XAI aims to assist regulatory audits, detect errors, and ideally, support stakeholders of algorithmic models in comprehending and interpreting the predictions of a system.

The prevailing moral literature predominantly situates XAI within a universal perspective, aligning its explainability with a fundamental right extended to stakeholders~\citep{DBLP:journals/csur/GuidottiMRTGP19, BARREDOARRIETA202082}. However, this perspective seems to encapsulate only a fraction of XAI's moral dimensions. 
To fully appreciate the ethical importance of XAI, we must venture beyond viewing it as a universal right, and consider it in terms of moral duty---a binding obligation to develop and implement XAI technologies that help ensure we can meet our established ethical duties and maintain an essential reciprocal relationship with human users.

This perspective resonates with the claim of Simone Weil that:

\blockquote{
\textit{The notion of obligation comes before that of rights, which is subordinate and relative to the former. A right is not effectual by itself, but only in relation to the obligation to which it corresponds}} (\cite{Weil1952}

We argue that we must consider XAI in terms of moral duty, or obligations, which take precedence over rights and form the bedrock upon which rights are upheld, if we are to capture its complete moral picture. To do so, we first acknowledge that establishing a cohesive definition of XAI proves challenging, given the diverging views held by various communities.

In the machine learning research sphere, XAI is perceived as a toolkit comprising technical methods \citep{BARREDOARRIETA202082} that illuminate crucial features in opaque models, akin to a \emph{reason explanation}~\citep{baum2022responsibility}. In contrast, policymakers regard it as a mechanism for auditing algorithms \citep{BellSNS22}. This divergence, highlighted in literature \citep{EdwardsV18, NanniniBS23}, calls for efforts to solidify explainability as a concept embraced universally across fields including machine learning, human-computer interaction, and governance.

The discourse surrounding XAI often intersects with the 'right to an explanation' stipulated in the  European Union's General Data Protection Regulation (EU GDPR) \citep{gdpr, ronan2021gdpr, watcherabs}, albeit revealing a disconnect with legislative knowledge and feasibility \citep{ronan2021gdpr}. Nonetheless, this association exemplifies the perceived prominence of explainability as a right, despite its surrounding ambiguities \citep{HackerP20, ebersXAIeu2022}.

XAI undoubtedly promises to aid stakeholders in deciphering AI predictions and identifying errors like model biases, thus establishing connections to relevant standards. Yet, considerable challenges loom, particularly in creating explanations that resonate with a diverse user base. Developers grapple with hurdles in integrating explainability methods \citep{BhattXSWTJGPME20}, while end-users demonstrate biases in interpreting these explanations \citep{BertrandBEM22}. The terminological discrepancies noted in policy documents further underscore a lack of comprehensive understanding of the nascent and intricate realm of explainability as a research concept \citep{NanniniBS23}.

Philosophically, while strides have been made in the empirical ethics of delineating XAI duties, particularly within the domain of public financial institutions~\citep{DBLP:conf/aies/Brand23}, a universal grounding of moral duties of XAI remains absent.

At the heart of this discussion, we propose that the ethical grounding of XAI should not merely be perceived as a universal right, but should also be intricately connected to moral duties within the context of AI. This thesis draws upon the notion that moral duties are reciprocally legislated between constituitively similar beings, e.g.~\cite{korsgaard1996norms, darwall2009second,Brand23, korsgaard2018fellow}.

This notion is particularly important as we aim to establish XAI as helping maintain this reciprocity in morally relevant relationship and contexts, signifying that it may inherently possess qualities that facilitate fair and just interactions, establishing a foundational platform for ethical discourse and engagement.

Applying this theoretical framework, we will assess whether current XAI capabilities align with the proposed ethical mandates. Connecting philosophical insights and practical observations will enable a nuanced discussion about moral duties.

\section{Reciprocity and Moral Duty}
\label{sec:2}
\subsection{Reciprocity as a Foundation for Moral Duties}
To examine the importance of XAI within the context of moral duties we must first consider the necessary conditions to sustain any moral duty. 

Briefly, moral duties refer to obligations, determined by underlying moral principles, that an individual, or group of individuals, has toward another individual, or group of individuals. A moral duty therefore forms an interpersonal relationship and is usually composed of a duty-holder with a directed duty to provide a certain action or service to another person, the claim-right holder. In other terms, whenever an individual has a claim-right to receive some service or treatment, they are making a claim that another individual has a duty to perform (or not) the action. As \cite{SimonCabulea2015} says, “\textit{Directed duties and claim-rights are, in effect, the same relation viewed from different perspectives.}” In this ethically-charged relationship there is someone whose claim or expectation of treatment---such as an accused claiming their right to a fair trial---entails a correlative directed duty for someone else to provide such right---such as the judge who must oversee that the trial of the accused follows the correct procedure~\citep{watson2021right}\footnote{While all claim-rights have a corresponding duty, it does not necessarily follow that all duties are directed to a corresponding claim-right, see for example \cite{watson2021right}. We set aside this case and solely consider the duty-rights relationship.}.

These two individuals, the duty-holder and right-holder, perceive the duty that grounds their moral relationship as reciprocal---by this we mean that to legislate any duty, whether it be the duty to respect others or the duty to provide truthful information, involves two persons, the right-holder and duty-holder, reciprocally communicating their position to each other. To explain this principle we turn to philosopher Christine Korsgaard. 

In contrast to Hume's account that moral duties are essentially social contracts based on power relations~\citep{hume1998enquiry}, Korsgaard stresses from the Kantian perspective that if a duty is understood as objectively true in any circumstance, it must have authority over all of those involved---the right-holder and the duty-holder and anyone who may later find themselves in either position. To illustrate what she means, Korsgaard provides the example of French laws: an individual can oblige another individual to act a certain way by using the authority of French laws. But this is only a legitimate claim on the person \emph{if and only if} they are French. The law must apply to them. And any person claiming that another person must adhere to this duty (or law) insofar as providing them with a reason to acknowledge and respect it, \emph{supposes that they too are bound by the same authority}~\citep{korsgaard2018fellow}. 

A small business owner, for example, can show up at a bank to ask for a loan and demand that the bank ought to review their application solely on merit and not on discriminatory factors; the bank and their employees thus have the correlated duty to respect the applicant and treat their application fairly. For this duty to be considered objective and not the mere subjective preference of the small business owner, they must also concede that if the roles were reversed, such that the they become the bank employee and the bank employee becomes the applicant, the bank employee (now applicant) could make the same claim. Moral duties are not \emph{unidirectional} insofar that one individual is recognizing and respecting another. Rather, they have a \emph{shared and reciprocal authority} where it is conceded that any rational being, regardless of the role in which they find themselves, would be equally committed to the claims and reasons of this moral law. Participants in this morally dutiful relationship must have this reciprocal, conceptual role-reversal for it to be considered as an objectively authoritative moral law.

\subsection{Reciprocity as Shared Authority} This notion of shared authority is crucial for moral duties. Shared authority implies that both the duty-holder and the rights-holder recognize the position of the other and share the reasons that place them equally under the same moral law.

Given this notion of shared authority, reciprocity also entails that both entities under the law be constitutively equal, otherwise they would not be able to conceptually reverse roles as required. AI, irrespective of the possibility that we may find sufficient technological advancements that allow AI to superficially replicate rational capabilities of a human, it will never be constitutively equal with humans. Primarily, machines do not justify actions with reference to moral norms the same way humans do (212–236p~\cite{asaro2020autonomous}). They also do not face the problem of vulnerability, such as concerns for mortality or emotions like insecurity, that is inherent to the human condition and which certainly shapes how we rationalize our conduct. Any machine which merely replicates human capabilities cannot switch roles from, for example, bank employee to loan applicant, the same way humans can. 

Two necessary requirements thus follow for any perceived moral duties wherein all subjects are placed under a reciprocally shared authority. First, both subjects, the duty-holder and rights-holder, participate in this relationship by actively recognizing the position of the other and sharing the reasons that place them equally under the authority of the moral law. They are both active \emph{reason-sharing agents}. They can also be a collective of individuals on either side, who can claim the duty or right based on their social position. 

It also follows that as each subject under the reciprocally shared authority conceptually have the possibility to switch roles, they must also share a sort of constitutive symmetry. For both individuals to conceive of themselves under the same authority, that they may find themselves in the opposing role, either right-holder or duty-holder, implies that they could indeed share the roles---therefore, it implies they must share basic constitutive qualities. Referring to the banking example above, but would apply to any moral duties involving humans, is the basic quality of being a human being. Lacking the qualities constitutive of all human beings, AI will always be in an asymmetrical relationship with humans and therefore, will never be able to participate in the reciprocal legislation\footnote{Korsgaard does rightly point out that in marginal cases, such as with children or other animals, reciprocal legislation of moral duties does not work \citep{korsgaard2018fellow}. With AI, however, it is always in the position of replacing fully rational human agents, such as a bank employee or legal professional. AI in decision-making context fills an authoritative role. The concern for marginal cases does not apply here.} underlying all moral duties \citep{Brand23}.

Since AI cannot participate in moral duties with humans, AI-assisted systems are not moral agents, despite appearing to make human-like decisions. If so, one may conclude these systems are \emph{amoral} and not bound to reciprocal legislation.
This, however, only takes into account one way of defining moral agency. While AI cannot exercise agency in the human-sense, we can also define agency in another way in which AI does fit, what James Moor calls \emph{implicit ethical agency}: when AI autonomously functions like human agents would in a specific moral context \citep{Moor2006machinethics}. AI-assisted systems fit within this role as it is deployed to make predictions, or decisions, in a bank for loan applications or making medical diagnoses for example. AI and humans are both, to different degrees, moral agents. Yet, given the difference of properties and capabilities between AI and humans, they lack the necessary reciprocal nature and, for the aforementioned reasons, cannot co-exist in the same moral setting.

\section{Reciprocity and XAI}

In Section \ref{sec:2}, we concluded that AI systems cannot possess moral agency and participate in reciprocal relationships with humans due to the lack of constitutive symmetry and shared rational capacities. This does not, however, preclude evaluating whether XAI tools can support human moral agency in contexts where AI systems are deployed to assist with decision-making.

While AI itself lacks moral standing, human operators utilizing AI predictions to help inform their decisions nevertheless should adhere to the moral norms of agency and reciprocity. XAI emerges as an interface to help sustain this moral situation by providing transparency into otherwise opaque AI systems. Explanations serve to empower human operators to genuinely understand and engage in the reasons supporting algorithmic predictions, rather than blindly follow them. This understanding supports meaningful human endorsement (or rejection) of AI suggestions.

In this sense, the moral assessment of XAI should focus on its capacity to maintain reciprocal human relationships despite the involvement of non-reciprocal AI systems. Though AI cannot strictly share moral duties with humans, XAI tools have instrumental moral value to the extent they help human operators understand and can genuinely interact with the decision-process as an interlocutor, therefore fulfilling their duties of agency and reciprocity. The goal is not to attribute reciprocity to AI itself, but evaluating whether XAI helps sustain reciprocity with humans interacting within an AI context.

With this framing linking the philosophical foundations to the XAI context, this Section examines whether current XAI capabilities and limitations are fulfilling this role of supporting human moral agency and reciprocity when relying on AI systems for decision-support.

\label{sec:3}
\subsection{Aligning XAI with Reciprocity}
Explainability of AI systems (XAI) has emerged as a critical interface between the algorithmic decision-making processes and the human moral and legal frameworks that govern such decisions. XAI operates through two primary paradigms: intrinsically interpretable models, often referred to as \enquote{glass-box} models, and post-hoc explanations that are applied to more complex, opaque models. The latter includes techniques like LIME~\citep{ribeiro2016should} and SHAP~\citep{lundberg2020local2global}, which aim to elucidate the decision-making process by highlighting key features and decision pathways \citep{BARREDOARRIETA202082}.


It also morally aligns with the fundamental concept of reciprocity. When someone uses AI as a decision-support system and such system is a machine-learning model, which is increasingly the case,
they are using the opaque \enquote{black box} model and thus, they are unsure of the reasons that justified the algorithmic decision. Endorsing an AI-produced recommendation in this opaque state is effectively making an arbitrary decision as the operator is no longer acting freely as an autonomous agent and instead is acting as a tool, or extension, of the AI itself. Without knowledge of the reasons supporting the decisions they endorse or reject, human operators are not genuinely engaging in reflective endorsement that is a necessary condition of human agency.
\subsection{Agency Requirement}
We take the intuitive position that agency includes an individual to initiate the action through their will, e.g.~\cite{396f9543-1bb7-3790-8b2c-7a2d6e7c3fa4}. Humans express their agency where they are not merely a vessel for an assortment  of natural inclinations/impulses, but when we are, “\textit{thinkers of our thoughts and the originators of our actions.}” (p. 377~\cite{korsgaard1996creating}). A human acting under her own agency does so by knowing and adopting reasons and principles that determine the course of her actions. Other animals act solely through instinct. For example, a person exercising their agency does not merely give money to a friend, but more specifically performs the action because they see their friend in need and recognize the importance of helping friends when in need. The bank employee does not offer loans at whim, but does so by examining the application, such that the applicant has a good income to debt ratio, and then considers these good reasons to give them the loan because they fit their principle of only offering loans to those who have the capacity to pay them back. Therefore when someone endorses the advice of an opaque AI decision-support system without knowing the principles or reasons supporting its decision-making process, they are acting against their autonomous agency.

As we argued above, moral duties not only necessitate constitutive symmetry between the duty-holder and right-holder to achieve the conceptual role-reversibility, but also requires them to be reason-sharing agents. Therefore if human operators approve decisions produced by opaque AI-systems without knowing the reasons that supported the decision, they are not genuinely exercising their agency, but are acting as instruments of the AI-system.\footnote{AI systems are in the case of decision-support systems implicit agents \citep{Moor2006machinethics}.} In this case, the only reason they could provide for endorsement is that \textit{the AI said so}, which suggests over-reliance and automation bias. The reasons in this context are superficial and not based on knowing the justificatory data.

XAI therefore has moral value in this context as its explanations \textit{help human operators understand} the decision-making process of the AI-system. By providing the operators with explanations of the data that supported the AI prediction, or \emph{motivating reasons} it took\footnote{\cite{baum2022responsibility} argue that XAI, specifically feature importance models, provides explanations akin to reason explanations.}, the operators can more genuinely interact with the system as an interlocutor, consider if the reasons are good reasons that align with their principles, and adopt them as their own reasons to endorse the decision. If they consider them as bad reasons, they can then reject the AI-predictions by contrasting it with what they consider to be good reasons that better match their principles. While XAI does not itself have normative character, its explanation are morally valuable to ensure human operators can maintain a genuine sense of agency essential for the reciprocal nature of morally dutiful relationships.

\begin{figure}[ht]
\centering
\includegraphics[width=0.95\textwidth]{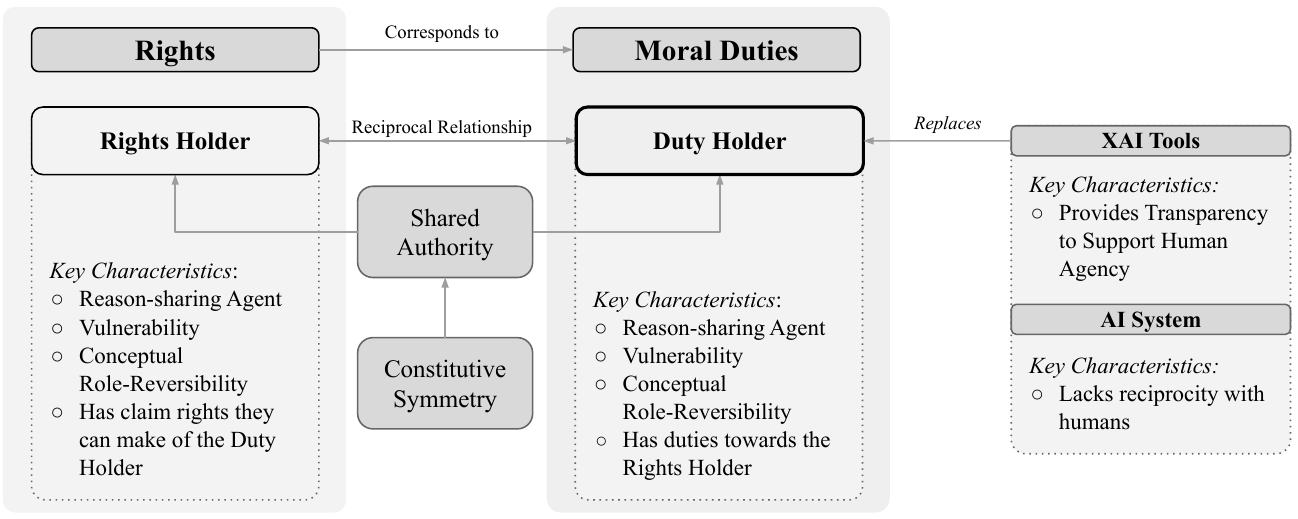}
\caption{A flow chart depicting the key concepts and relationships from the philosophical framework in Section \ref{sec:2} \& \ref{sec:3}.}
\label{fig:flowchart}
\end{figure}

The flow chart in Figure \ref{fig:flowchart} condenses the framework detailed in this section into a visual summary. Understanding the moral potential of XAI to ensure the sustaining of our moral duties, the next section will examine the current approaches in various communities to XAI research, development, and implementation. By doing so, this will identify the limits of XAI to better understand how it can support reciprocity and our moral duties to one another.

\section{Practical Limits of XAI: The Multidisciplinary Challenge of Actualizing Reciprocity}
\label{sec:4}

Reciprocity mandates XAI tools that provide transparent and accurate explanations of opaque AI-systems to empower human operators and help them maintain their agentic control. However, the efficacy of XAI in fulfilling this role is contingent upon its ability to offer truthful representations of the algorithmic decision process. This is an area still under active research and debate, particularly concerning the limitations of post-hoc XAI methods \citep{rudin2019stop, Zerilli2022}.

Given the complex landscape of XAI, which spans multiple disciplines including machine learning, human-computer interaction, and AI policy governance, accomplishing the requirements of reciprocity presents a multifaceted challenge. Each of these communities confronts unique barriers, from practical limitations to ethical oversights. This section shifts our philosophical analysis to the practical by synthesizing these diverse perspectives to offer a comprehensive understanding of the obstacles and opportunities in developing XAI systems to support human moral agency and moral duties.

\subsection{Machine Learning Community}
\paragraph{Technical Methods in XAI --}
    Within the ML sphere, XAI refers to an assortment of techniques and algorithms that aim to account for the behavior and predictions of complex algorithms. 
Explainability techniques are increasingly used to overcome the difficulty of interpreting AI decisions, with techniques such as feature importance attribution~\citep{NIPS2017_7062,lundberg2020local2global}, algorithmic recourse~\citep{karimi2021algorithmic}, model copying \citep{unceta2020copying} or knowledge distillation~\citep{gou2021knowledge}. 
Recent research has focused on developing model-agnostic XAI techniques to explain any opaque ML model. The emphasis is on mathematical and algorithmic techniques that aim to elucidate the inner workings of the model and attribute importance to inputs. Explanations aim to provide technical insight into model behavior, though may lack resonance with non-expert stakeholders~\citep{BhattXSWTJGPME20,DBLP:journals/cacm/Lipton18}.


\paragraph{Real-world Effectiveness and Limitations --}
%
Yet, while the ML community has developed an extensive toolkit of technical methods aimed at explaining opaque models, the real-world effectiveness and adoption of these techniques remains questionable. Qualitative studies have found that providing end users with XAI explanations does not necessarily improve their understanding, trust or capacity to leverage AI systems \citep{WangY21Helpful, KaurNJCWV21, LiaoZLDD22, BellSNS22}. The mathematical explanations generated often lack resonance with non-expert stakeholders. Moreover, many techniques rely on assumptions that the AI model faithfully represents true underlying relationships, which is not always the case.

Organizational constraints around data privacy, integration, and compliance also impose barriers to implementing XAI tools in applied settings \citep{BansalWZFNKRW21, Fok2023bs-2305-07722}. The ML community often focuses narrowly on algorithmic advances without fully considering the broader AI governance context, such as data integration, intellectual property, and model systems~\citep{SambasivanKHAPA21, GeorgievaLTV22, IbanezO22}. Additionally, technical XAI approaches frequently fail to account for diversity in users' backgrounds, intentions, and decision-making processes. Consequently, meaningful explanations for experts are not necessarily intuitive explanations for other users.

%
%
%

\subsection{Human-Computer Interaction (HCI) Community}
\paragraph{Human-Centric Aspects of XAI --}
Within HCI, XAI research concentrates more heavily on the human side of the equation, specifically, how explanations affect trust, transparency, and the capacity of users to sufficiently understand AI systems. HCI experts conduct user studies analyzing how people interact with and comprehend different explanation interfaces, formats, and contents. Researchers explore qualities of explanations that make them useful, interpretable, and meaningful for diverse audiences \citep{explainindecisiona2019mao, WangYAL19}. With XAI, HCI emphasizes the importance of generating explanations that relate to users' mental models and background knowledge so they can critically evaluate AI behaviors. Nonetheless, HCI research can overlook technical considerations of explanation methods and thus favour explanations with intuitive understanding that do not necessarily capture a model's true logic. Although, recent work questions the need for a comprehensive understanding of an algorithm to sufficiently make justifiable decisions \citep{Zerilli2022,nazar2021systematic}.

\paragraph{Reciprocity and Moral Agency --}
The human-centred lens of the HCI community supports reciprocity through XAI design by focusing on explanations that match users' mental models and backgrounds. HCI researchers aim to provide not only transparent but accessible and intelligible explanations for users. HCI, however, must nevertheless balance accessibility with fidelity to the system to support informed consent and confidence in the system's capabilities \citep{VasconcelosJGGBK23}. This requires collaboration with domain experts to design explanations that accurately convey model limitations, uncertainties, and flaws. It also requires a longitudinal study of whether simplified explanations sustain informed trust from users as systems evolve in time. It is therefore preferential that HCI research takes advantage of interdisciplinary expertise to create explanations that are not just faithful representations of the system, but are also genuinely understood by different stakeholders \citep{TalKK22, LiaoGM20}.

\paragraph{Limitations and Broader Implications --}
However, while HCI studies offer invaluable insight into how explanations affect trust and transparency, their controlled experiments do not always take into account real-world complexities that shape XAI's capacity to enable genuine interaction with the system \citep{HolsteinVDDW19}. HCI's simplified lab settings with compensated participants often fail to reflect natural organizational contexts \citep{IbanezO22, GeorgievaLTV22}. Indeed, focusing solely on understanding explanations neglects existing technical and social factors critical to then take action informed by them \citep{Robbins19, LiaoZLDD22}.

If explanations arise from inaccurate models or mismatch user expertise levels, the transparency provided by XAI is fundamentally limited. And common organizational constraints around data, privacy, compliance, and power frequently mitigate practical usefulness of XAI methods  \citep{BansalWZFNKRW21, Fok2023bs-2305-07722}. While HCI provides preliminary insight into the prospects of XAI helping provide meaningful explanations to users and thus, aid in sustaining reciprocity, it is important to acknowledge socio-technical constraints and demands that also effect how explanations are provided to users. 
%
Ultimately, the controlled focus of HCI offers an important starting point to explore the connection between the requirements of reciprocally legislating moral duties and XAI, but must eventually be contextualized within broader technical and social dynamics to determine the capacity of XAI in this moral context.

\subsection{Policymakers}
\paragraph{Regulatory Aims and XAI --}For policymakers and regulatory bodies, XAI serves as a mechanism for auditing algorithms, detecting errors or biases, and evaluating compliance with legal standards around transparency and accountability~\citep{HackerP20, ebersXAIeu2022, NanniniBS23, DBLP:conf/pkdd/NavarroKG21}.
The aim is to establish a legal relationship between AI systems and users by ensuring transparency and accountability.
%
XAI could provide oversight into a variety of moral and legally salient automated decision-making processes like access to loans, social services, and criminal justice outcomes. XAI helps enable scrutiny of whether algorithmic systems align with fairness and non-discrimination laws. Policymakers therefore emphasize the role of XAI in accountability, recourse, and due process in contexts of moral and legally salient AI applications.

\paragraph{Reliance and Ambiguity --}
%
While policymakers promote XAI as an oversight tool for compliant AI, they often rely on ML and HCI conceptualizations without independent technical assessment \citep{BellSNS22}. This reliance can undermine the goal of receiving adequate assessments of the motivating reasons used by a system. 
%
Consequently, XAI definitions in policies frequently remain ambiguous, failing to delineate substantive thresholds for sufficiency of explanations --- i.e., what target and typology of explanation shall be provided. This leads to situations of legal uncertainty, where AI providers might be required to provide explanations of generic informativeness. This translate to the risk of not being able to genuinely evaluate model behaviors, as well as the risk of ethics-washing \citep{NanniniBS23}. The absence of clear criteria for what makes an explanation legally or ethically \enquote{sufficient} creates a gap in the legislation, affecting the potential for a robust system of explanations \citep{ebersXAIeu2022}.

\paragraph{Case Studies and Practical Challenges --}
%
Moreover, policies that impose XAI requirements commonly overlook feasibility constraints related to data privacy, model complexity, and trade secrecy. Mandating local explanations for any AI system regardless of proprietary or security issues risks unrealistic regulations. Further illustrating inadequate technical grounding, some policy documents have exhibited fundamental misunderstandings over definitions and properties of AI explainability methods\footnote{Regarding definitions, we refer to the case of the EU's \cite{EthicsGuidelinesTrustworthy2018} that published the \textit{Ethics guidelines for trustworthy AI}. The document presented AI explainability within the term \enquote{explicability}. The guidelines were consulted by the European Commission for the first draft release of their AI Act release in November 2021 \citep{aiactoriginal}. Following amendments offered glances of alternated terminological choices and properties - oftentimes with connotations of XAI redirected
more towards system oversight rather than end-users' service. This heterogeneity is witnessed in the draft versions released by the committees, such in the IMCO-LIBE's amendments voted and approved ahead of the 'Trilogue' phase in June 2023 by the European Parliament \citep{aiactgeneralapproach, aiacttrilogues}}.
%
For instance, debate continues around the \textit{GDPR}'s \emph{right to explanation} as it lacks precise criteria for assessing when provided explanations sufficiently describe the reasoning logic behind a given automated decision \citep{gdpr, malgieriWhyRightLegibility2017, wachterWhyRightExplanation2017,  ebersXAIeu2022}. This ambiguity hinders the practical implementation of XAI if users are unable to collectively assess and agree whether any given AI system and their explanations are indeed what is expected, a \enquote{sufficient explanation.}

\section{Discussion}
The examination of leading XAI research communities reveals gaps between the proposed philosophical reciprocity framework and practical constraints. While aspirations exist to develop XAI that upholds moral considerations, each community currently falls short in key aspects.
Progress in the practical ethical implementation of XAI requires bridging these gaps through creative interdisciplinary collaboration. We hereby synthesizes opportunities and responsibilities for communities to collectively advance XAI implementation that supports our moral objective.

\paragraph{ML Community: Bridging the Gap Between Novelty and Applicability --}The ML emphasis on novel techniques often overshadows the necessity for real-world evaluation. While linear model approximations may serve as a convenient heuristic, they often fall short in conveying the intricacies of complex neural networks to lay users. To address this, a paradigm shift towards "Collaborative XAI Design" is warranted. This would involve co-creation workshops that bring together domain experts, ethicists, and end-users to collaboratively identify the cognitive and contextual gaps that hinder effective intelligibility and reason-sharing \citep{LiaoZLDD22}. Longitudinal studies could then be conducted to assess the efficacy of these collaboratively-designed XAI systems in how they enable or inhibit user understanding across evolving AI behaviors \citep{CabitzaCMNSSH23}. Additionally, research should focus on identifying and overcoming barriers to the organizational deployment of such systems, such as resistance to change or lack of technical expertise within the organization \citep{IbanezO22, GeorgievaLTV22}.

\paragraph{Human-Computer Interaction: Beyond Simplified Lab Studies --} While HCI research provides valuable insights into user understanding and interaction, it often falls short in capturing the complexities of real-world settings. Simplified lab studies, for instance, are unable to account for the influence of power dynamics and organizational incentives that may lead users to uncritically accept AI explanations. To address this, ethnographic research methods should be employed to capture the nuances of organizational culture and behavior over extended periods. These studies may reveal tacit barriers to the effective provision and utilization of explanations, such as hierarchical structures that discourage questioning or incentives that prioritize rapid deployment over thorough evaluation. Collaboration with subject matter experts in fields like psychology and organizational behavior could further enrich these studies by providing frameworks for assessing the accuracy and comprehensibility of AI explanations, thereby minimizing the risks of misplaced user trust due to oversimplification \citep{VasconcelosJGGBK23, LiaoZLDD22, Bell2023abs-2207-01482}. 

\paragraph{Policy and Governance: Grounding Reciprocity in Technical Realities --} Current policy-making efforts in the realm of XAI often suffer from a lack of technical understanding, leading to regulations that are either overly ambitious or woefully inadequate. To address this, multidisciplinary efforts involving ML researchers, ethicists, legal experts, and policymakers should be convened to discuss the feasibility and effectiveness of proposed XAI regulations. Encouraging signs of such initiatives are arising \citep{projectexplain, ExplainingDecisionsMade2020, NIST-8312, NIST-8367, NISTxaiWG, AlonsoBBDGGRSTT20}, yet still foremost grounded on national AI policy studies or workshops rather than sustained and cross-institutional initatives. Independent testing initiatives should be established to assess XAI systems against a set of formalized criteria based on reciprocity, such as relevance, completeness, and potential impact on human rights. These criteria should be dynamic, evolving in response to advancements in XAI capabilities and methodologies \citep{NanniniBS23, BellSNS22, Bell2023abs-2207-01482}. This would allow for the development of adaptive standards that can tie ambitious regulatory goals to achievable, reciprocity-focused oversight mechanisms.

\section{Conclusion}

In this paper we explored XAI by elevating it from a mere technological or regulatory construct to a morally significant tool, rooted in facilitating the principle of reciprocity, an essential concept to moral duties. In doing so, it addresses a notable gap in existing XAI ethical literature, which has largely framed XAI as a universal right, thereby neglecting broader moral dimensions. By showing how XAI can serve as a pivotal mechanism for fulfilling our moral duties in the context of AI-systems, we show how it thereby forms the foundational moral basis upon which rights, such as a right to explanations, are meaningfully upheld.
By examining the perspectives of key stakeholder communities, we have illuminated that while each has made valuable contributions, they often operate in isolation, lacking a unified ethical framework. This absence of a cohesive ethical lens poses a significant challenge to implementing the moral ideal presented in Section \ref{sec:2} and \ref{sec:3} across diverse domains.

In summary, this paper fills a critical gap in existing XAI ethical literature by shifting the focus from rights to an emphasis on moral duties and reciprocity. This provides a foundational platform for ethical discourse and engagement. While challenges in real-world efficacy and implementation persist, this paper provides a comprehensive ethical framework through which the development and implementation of XAI can be more meaningfully guided.

\acksection

The research and PhD for J. Brand is funded by the XAI4AML chair (N° ANR-20-CHIA-0023-01) and supported by the OpAIE team at Télécom Paris, Dr Winston Maxwell, and Dr John Zerilli.
L. Nannini received funding contribution from the ITN project NL4XAI (\textit{Natural Language for Explainable AI}). This project has received funding from the European Union’s Horizon 2020 research and innovation programme under the Marie Skłodowska-Curie grant agreement No 860621. This document reflects the views of the author(s) and does not necessarily reflect the views or policy of the European Commission. The REA cannot be held responsible for any use that may be made of the information this document contains.

The authors wish to personally acknowledge Carlos Mougan (Southampton University, UK) for the thoughtful feedback.

\bibliography{references}
\bibliographystyle{apalike}

\end{document}